# Secure Steganography Technique Based on Bitplane Indexes

Alan Anwer Abdulla, Sabah A. Jassim and Harin Sellahewa
Applied Computing dept.
University of Buckingham
Buckingham, UK
alananwer@yahoo.com {sabah.jassim; harin.sellahewa}@buckingham.ac.uk

*Abstract*— **This paper is concerned with secret hiding in multiple image bitplanes for increased security without undermining capacity. A secure steganographic algorithm based on bitplanes index manipulation is proposed. The index manipulation is confined to the first two Least Significant Bits of the cover image. The proposed algorithm has the property of un-detectability with respect to stego quality and payload capacity. Experimental results demonstrate that the proposed technique is secure against statistical attacks such as pair of value (PoV), Weighted Stego steganalyser (WS), and Multi Bitplane Weighted Stego steganalyser (MLSB-WS).**

## I. INTRODUCTION

With the rapid growth in Internet transactions over fast insecure communication channels, the security and confidentiality of transmitted sensitive data have become a serious concern in information communication. In comparison to cryptography, steganography secures communication between two parties by hiding the very existence of the secret message rather than scrambling it. Steganography embeds sensitive data (e.g., secret messages, images, audio files) in unsuspecting cover media such as images, audio and video so that attackers or unauthorized persons do not realize that sensitive data are being transmitted. The media that contains the secret message is called a stego. The focus of this paper is on embedding secret bits in images.

Existing methods used in image steganography can be grouped into two main categories based on the hiding domain: spatial domain and frequency domain methods. Spatial domain methods typically embed message bits in a cover image by using least-significant-bit (LSB) replacement techniques. Location of embedding cover pixels could be selected sequentially [1,2,3], randomly [3] or edge based [4]. Frequency domain techniques first transform the image into a frequency domain such as DCT [5] or wavelet [6] and embeds the secret by adjusting the coefficients in certain frequency ranges. Our proposed method is a spatial domain embedding scheme and is closely related to LSB based techniques.

LSB steganography is based on manipulating the LSB bitplane by directly replacing the LSBs of cover-image pixels with the message bits [1,2,3]. The advantage of these methods is the relatively high stego quality and payload capacity, since only minor changes are made to cover pixel intensity and each cover pixel is used to carry a secret bit. However the secret message can be detected easily, by steganalyser techniques, which we will discuss shortly. In [7], the payload capacity of the most-LSB replacement is improved by replacing the first two LSBs (2LSB) with 2 secret bits, i.e., 2 secret bits can be embedded per each cover pixel. However, embedding two bits per pixel increases the changes introduced to the cover image adversely affecting stego quality, making the stego easier to detect. In [8], a technique based on mapping/embedding two secret bits in the first three Fibonacci LSBs of the cover image pixel is proposed. This technique enhances payload capacity and un-detectability at the cost of reduced stego quality. Other techniques have been produced to improve stego quality and un-detectability by reducing payload capacity [9,10].

Whilst steganographers aim to design techniques to increase capacity and robustness of stegos and introduce minimal changes to the cover, steganalysers attempt to defeat the goal of steganography by detecting the presence of a hidden message. There are a number of statistical attacks to determine the presence/absence of a hidden message and estimate the size of the embedded secret message. Three well-known and reliable steganalysis techniques extensively used to detect and estimate the secret message are PoV [11], WS steganalyser [12], and MLSB-WS [13].

Some steganographic techniques attempt to be robust against steganalysis, but such schemes usually have payload capacity limitation. Decreasing the payload capacity increases the un-detectability [9,10]. It is not easy for a steganographer to achieve a good balance among the different steganography requirements such as payload capacity, un-detectability, stego quality, and robustness against active attacks.

Our proposed steganographic algorithm aims to provide robustness against detection by the statistical steganalysis attacks described in [11,12,13], whilst achieving an acceptable level of payload capacity and stego quality.

The rest of the paper is organized as follows: the proposed technique is presented in section II. The methodology adopted to evaluate the proposed technique is described in Section III. Section IV presents experimental results and the paper is concluded in section V.

## II. PROPOSED METHOD

In LSB randomly technique each pixel of the cover image can be used for embedding one secret bit with very small image degradation but it is easy to detect the secret bits by the statistical steganalysers, while 2LSB embedding technique has double the capacity of LSB, but quality of the stego image deteriorates more and the secret can be detected easily by statistical steganalysers. The proposed technique aims to maintain the capacity as LSB randomly, use the two LSB bits to embed but only in one of them at a time and thereby maintain better stego quality than the 2LSB scheme, and yet ensures robustness against statistical steganalyaser such as PoV, WS, and MLSB-WS.

## A. Embedding procedure

- The cover image is first pre-processed by modifying the 2LSBs of each pixel in the original image as follows:

$$2LSBs = \begin{cases} 01 & if\ 2LSBs = 11 \\ 10 & if\ 2LSBs = 00 \\ 2LSBs & otherwise \end{cases} \quad (1)$$

- One secret bit is embedded in each pixel. The secret bit is first compared with the first LSB of the modified cover pixel. If they are equal, then record the index of the first LSB plane. Otherwise, record the index of the second LSB plane (i.e., record 0 if the secret bit matches the first LSB; record 1 if the secret bit matches the second LSB).

- For the next secret bit, check the same similarity. This time the record value of the index must be different from the previous one because vector of indices must be in form of 10s or 01s, i.e. if the previous index value was 1, the next index value must be 0, otherwise swap the first two LSBs of the cover pixel.

- Finally, the vector of indices is either of the form 1 0 1 0….1 0  or 0 1 0 1 …. 0 1, i.e., each index value differs from the previous one. This vector must be sent to the receiver in a form n(10) or n(01), n is the number of repeating 10s or 01s in vector. For example if there are one thousand secret bits, then the receiver should get 500 (10) or 500 (01).

## B. Extracting procedure

- Depending on the n(10) / n(01) the receiver create the vector of indices.
- If the element of the vector of indices is 0, it means the secret bit must be extracted from the first LSB of the selected stego pixel, otherwise (i.e. the element is 1) the secret bit must be extracted from the second LSB. All bits can be extracted by repeating this procedure.

Example:

If we have the secret bits 0 0 1 0, and four cover pixels whose first two LSBs are 01, 01, 10, 10. The first secret bit (which is 0 here) is compared with the first LSB (which is 1) of the first selected cover pixel. Because they are not equal then we compare the secret bit with second LSB (which is 0), now they are equal and we record the index value 1 indicating that the secret bit is similar to the second LSB of the selected cover pixel. The next secret bit (which is 0) is compared with the first LSB of the next selected cover pixel (which is 1), because they are not equal then the secret bit must be compared with the second LSB (which is 0) and now they are equal but cannot record the index value 1 because the previous index value was 1, in this case do the swapping between the first and second LSBs, i.e. change 01 by 10, and now the secret bit is similar to the first LSB then record index value 0. Continuing in this way we get a vector of indices such (1, 0, 1, 0). Now the sender should send 2(10) to the receiver indicating 2 pairs of 10s. The receiver checks the first index value of the vector of indices, if it is 0 it means extract from the first LSB of the first selected stego pixel, otherwise extract from the second LSB, and repeats the same thing for the other extracting process.

## III. EVALUATION METHODOLOGY

### A. Data

For testing the proposed scheme, five different cover-images (see Fig. 2) of size 512 x 512 are used. The secret bits have been generated using the Matlab PRNG (Pseudo Random Number Generator).

### B. Quality

Generally, peak signal-to-noise ratio (PSNR) is most commonly used as a measure of the quality of the stego image in field of steganography [14,15]. A larger PSNR value means that the stego image preserves the original cover image quality better.

### C. Detectability

Three steganalysers are used to evaluate the proposed:

*1) (PoV) steganalyser [11]:* Is a test that makes the statistical probability of embedding using Chi-square test. This steganalyser is detecting the secret message that embedded in first LSB.

*2) WS steganalyser [12]:* This steganalyser is invented to detect and estimate the secret message length embedded in the LSB.

*3) MLSB-WS steganalyser[13]:* This steganalyser technique aims to detect and estimate the secret message length. The technique is able to detect each bitplane of the cover pixel separately i.e. this technique is designed for multi bitplanes embedding technique.

## IV. EXPERIMENTAL RESULTS

To evaluate the proposed technique, two experiments are conducted. The first is to test detectability and the other is to compute the stego quality. Then the results are compared with the two known steganographic techniques (LSB randomly [3], and 2LSB [7]). For each tested technique and for each cover image, we have generated five stego images by embedding five different message length p = 0.2, 0.4, 0.6, 0.8, and 1 corresponding to 20%, 40%, 60%, 80% and 100% of the total pixel number of the cover-image respectively. After embedding, for each case, i.e. for each technique with a specific embedding rate, 5 stegos are produced.

### A. Detectability

To evaluate robustness of the proposed technique against detectability, the following three steganalysers have been used to attack it, and compare the results with those for LSB randomly and 2LSB embedding techniques.

*1) Pair of value (PoV) steganalyser:*

When an image is taken through the *PoV*, a chart is output that plots the probability of embedding against the percentage of the image pixels tested positive. If the steganalyst is presented with a plot similar to that in Fig. 3 (A), then the image should be assumed to have not been manipulated, while if the plot is like (B) or (C) in Fig. 3, then the image is assumed to contain 50% or 100% of the secret respectively. For the convenience of display, the result of only three images, a, b and c (see Fig. 2), out of five images and only for 100% embedding capacity is displayed. From Fig. 7, we can notice that all plots of our proposed are exactly like Fig. 3 (A) which means that they are considered to be non-stego images, i.e. it is robust against PoV in contrast to LSB randomly and 2LSB embedding techniques both of which are not robust.

*2) WS steganalyser:*

When an image is submitted to this steganalyser, a real number is output which should indicate the probability of having a secret hidden. A negative value is treated as 0 and any number >1 is an indication of full 100% secret load.

Table I shows the estimation results of the secret message length of three steganographic techniques including our proposed one by this steganalyser [12]. From Table I part C, it is noticeable that the proposed scheme mostly scores a negative value indicating absence of embedded secret even when the images have a 100% load. On the other hand the LSB randomly and 2LSB technique (see part A and B in Table I respectively) the stegoanalyser detects the secret message with high probability.

*3) MLSB-WS steganalyser:*

Again when an image is submitted to this steganalyser, a real number is output which should indicate the probability of having a secret hidden. A negative value is treated as 0 and any number >1 is an indication of full 100% secret load.

Table II shows the results of the estimation of secret message length for each first (L=1) and second (L=2) LSB separately for 2LSB embedding technique and for our proposed technique. While only first LSB (L=1) of the LSB randomly technique has been detected because the secret message is embedded in only first LSB. From Table II part D, it is noticed again that the proposed technique robust against this steganalyser, while for the second LSB (i.e. part E in Table II) this steganalyser can detect and estimate the secret bit only when the embedding ratio is 100% otherwise cannot detected.

*B. Stego Quality*

The average PSNR values calculated between the covers and stegos for each technique are illustrated in Fig. 1. This shows that the PSNR of the proposed is very close to the PSNR of 2LSB and PSNR of both of them are lower than the PSNR of LSB randomly. This result can be attributed to the fact that for the LSB randomly only first LSB has changed; while in the proposed technique both of first and second LSB has changed and in 2LSB technique both of first and second LSBs are changed.

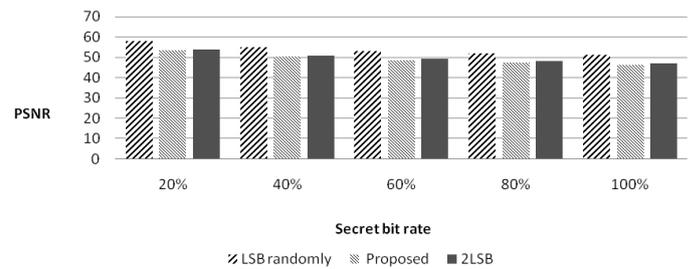

Figure 1. Stego quality

V. CONCLUSION

A secure steganographic algorithm based on bitplane indexes manipulation has been proposed. It is robust against detectability by three different steganalysers. This algorithm outperforms two well-known steganography schemes (LSB randomly and 2LSB) with respect to stego quality and payload capacity. In some way the proposed scheme is a hybrid of the LSB randomly and 2LSB but with improved un-detectability and good stego quality.

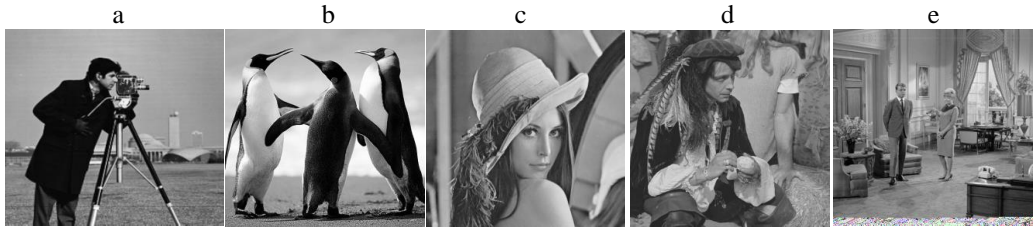

Figure 2. Cover images

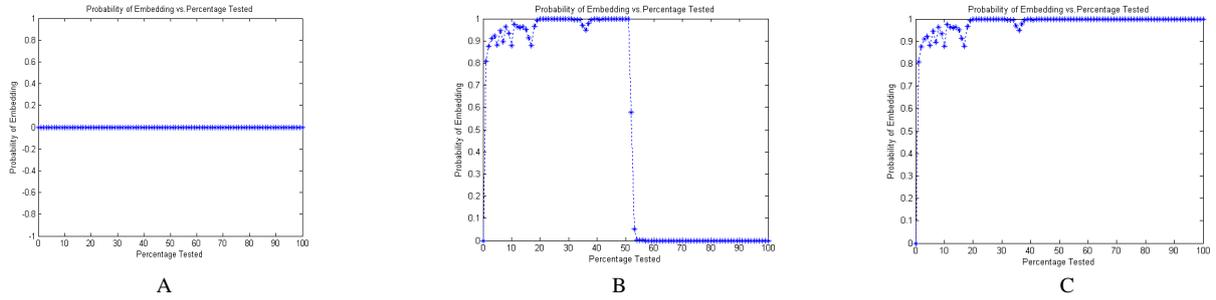

Figure 3. Examples of PoV plots: (A) zero embedding, (B) 50% embedding, and (C) 100% embedding

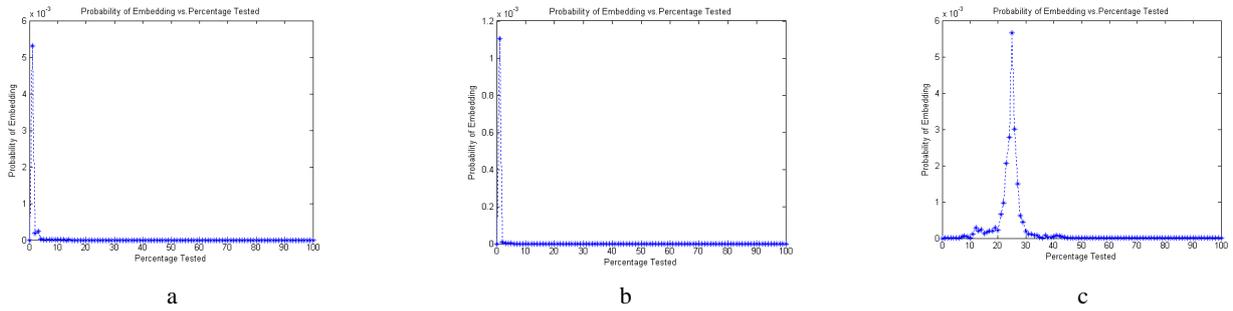

Figure 4. PoV of cover-images (i.e. no embedding)

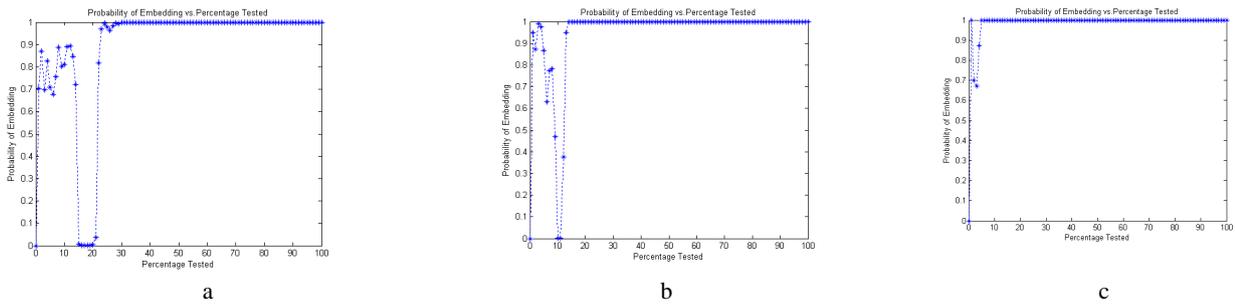

Figure 5. PoV of LSB randomly (100%) embedding

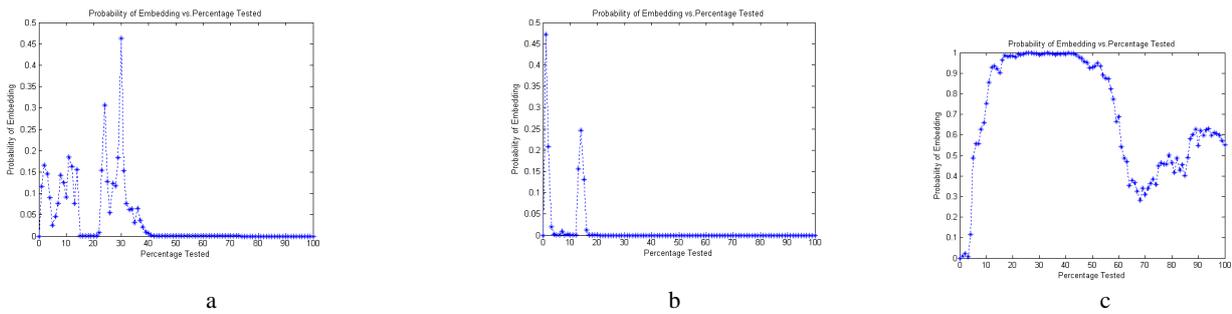

Figure 6. PoV of 2LSB randomly (100%) embedding

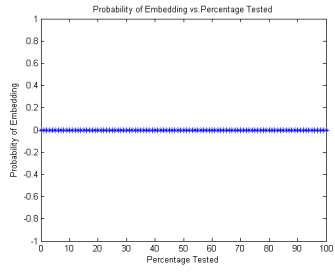 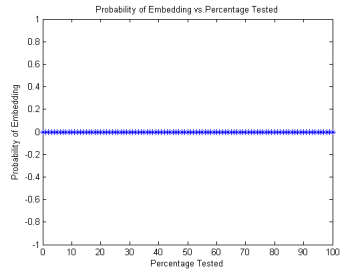 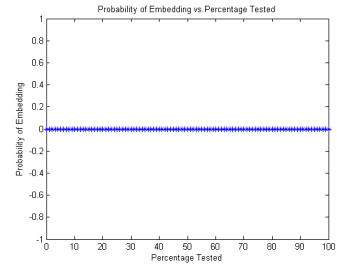

     a                              b                              c

Figure 7. PoV of Proposed (100%) embedding

TABLE I. STEGANALYSER [12]

**A. WS steganalyser on LSB randomly**

| images | Embedding ratio | | | | | |
|---|---|---|---|---|---|---|
| | *0 %* | *20 %* | *40 %* | *60 %* | *80 %* | *100%* |
| a | -0.017 | 0.174 | 0.364 | 0.551 | 0.757 | 0.976 |
| b | 1.095 | 1.049 | 1.039 | 0.986 | 0.953 | 0.926 |
| c | -0.017 | 0.188 | 0.393 | 0.574 | 0.772 | 0.992 |
| d | -0.023 | 0.192 | 0.435 | 0.601 | 0.823 | 1.026 |
| e | -0.025 | 0.202 | 0.425 | 0.609 | 0.826 | 1.022 |

**B. WS steganalyser on 2LSB**

| images | Embedding ratio | | | | | |
|---|---|---|---|---|---|---|
| | *0 %* | *20 %* | *40 %* | *60 %* | *80 %* | *100%* |
| a | -0.017 | 0.087 | 0.197 | 0.307 | 0.401 | 0.495 |
| b | 1.095 | 1.087 | 1.048 | 1.064 | 1.037 | 1.028 |
| c | -0.017 | 0.092 | 0.188 | 0.294 | 0.382 | 0.471 |
| d | -0.023 | 0.076 | 0.171 | 0.279 | 0.395 | 0.466 |
| e | -0.025 | 0.092 | 0.186 | 0.287 | 0.405 | 0.509 |

**C. WS steganalyser on Proposed**

| images | Embedding ratio | | | | | |
|---|---|---|---|---|---|---|
| | *0 %* | *20 %* | *40 %* | *60 %* | *80 %* | *100%* |
| a | -0.017 | -0.207 | -0.399 | -0.589 | -0.797 | -0.975 |
| b | 1.0955 | 0.6718 | 0.263 | -0.155 | -0.521 | -0.905 |
| c | -0.017 | -0.198 | -0.392 | -0.592 | -0.775 | -0.957 |
| d | -0.021 | -0.224 | -0.408 | -0.607 | -0.795 | -1.001 |
| e | -0.025 | -0.199 | -0.380 | -0.572 | -0.767 | -0.952 |

TABLE II. STRGANALYSER [13]

**A. MLSB-WS steganalyser on LSB randomly (L=1)**

| images | Embedding ratio | | | | | |
|---|---|---|---|---|---|---|
| | *0 %* | *20 %* | *40 %* | *60 %* | *80 %* | *100%* |
| a | -0.017 | 0.1747 | 0.3641 | 0.5513 | 0.7572 | 0.976 |
| b | 1.0955 | 1.0497 | 1.0393 | 0.9805 | 0.953 | 0.926 |
| c | -0.017 | 0.1886 | 0.3939 | 0.5745 | 0.7726 | 0.9927 |
| d | -0.023 | 0.1928 | 0.4357 | 0.6017 | 0.8236 | 1.0263 |
| e | -0.025 | 0.2023 | 0.4254 | 0.6091 | 0.8264 | 1.0227 |

**B. MLSB-WS steganalyser on 2LSB (L=1)**

| images | Embedding ratio | | | | | |
|---|---|---|---|---|---|---|
| | *0 %* | *20 %* | *40 %* | *60 %* | *80 %* | *100%* |
| a | -0.017 | 0.0877 | 0.1971 | 0.3006 | 0.4014 | 0.4953 |
| b | 1.0955 | 1.0809 | 1.0326 | 1.0385 | 1.0054 | 0.9893 |
| c | -0.017 | 0.092 | 0.188 | 0.2942 | 0.3822 | 0.4711 |
| d | -0.023 | 0.0758 | 0.1708 | 0.2795 | 0.3951 | 0.4661 |
| e | -0.025 | 0.0925 | 0.1868 | 0.2874 | 0.4005 | 0.5093 |

**C. MLSB-WS steganalyser on 2LSB (L=2)**

| images | Embedding ratio | | | | | |
|---|---|---|---|---|---|---|
| | *0 %* | *20 %* | *40 %* | *60 %* | *80 %* | *100%* |
| a | -0.017 | 0.0979 | 0.2021 | 0.3051 | 0.4066 | 0.5022 |
| b | 1.0955 | -0.1375 | -0.0066 | 0.1068 | 0.2354 | 0.3478 |
| c | -0.017 | 0.0977 | 0.202 | 0.3069 | 0.404 | 0.5003 |
| d | -0.021 | 0.0759 | 0.1771 | 0.2781 | 0.3864 | 0.4783 |
| e | -0.023 | 0.0784 | 0.18 | 0.2858 | 0.385 | 0.489 |

**D. MLSB-WS steganalyser on Proposed (L=1)**

| images | Embedding ratio | | | | | |
|---|---|---|---|---|---|---|
| | *0 %* | *20 %* | *40 %* | *60 %* | *80 %* | *100%* |
| a | -0.017 | -0.2077 | -0.3993 | -0.589 | -0.797 | -0.975 |
| b | 1.0955 | 0.6718 | 0.263 | -0.155 | -0.521 | -0.954 |
| c | -0.011 | -0.1989 | -0.392 | -0.592 | -0.775 | -0.950 |
| d | -0.021 | -0.2247 | -0.4008 | -0.607 | -0.795 | -1.001 |
| e | -0.053 | -0.1939 | -0.3803 | -0.572 | -0.763 | -0.952 |

**E. MLSB-WS steganalyser on Proposed (L=2)**

| images | Embedding ratio | | | | | |
|---|---|---|---|---|---|---|
| | *0 %* | *20 %* | *40 %* | *60 %* | *80 %* | *100%* |
| a | -0.009 | 0.0527 | 0.1336 | 0.2475 | 0.4631 | 0.9735 |
| b | -0.259 | -0.2129 | -0.108 | 0.0269 | 0.2361 | 0.9054 |
| c | 0.0008 | 0.0495 | 0.1263 | 0.2417 | 0.4296 | 0.9507 |
| d | -0.035 | 0.0314 | 0.1143 | 0.2093 | 0.4343 | 1.0021 |
| e | -0.020 | 0.0272 | 0.1049 | 0.2062 | 0.4214 | 0.9524 |